\documentclass[aip]{revtex4-1}
\usepackage[english]{babel}
\usepackage{graphicx}% Include figure files
\usepackage{dcolumn}% Align table columns on decimal point
\usepackage{bm}% bold math
\usepackage[mathlines]{lineno}% Enable numbering of text and display math
\usepackage[utf8]{inputenc}
\usepackage[T1]{fontenc}
\usepackage{mathptmx}
\usepackage{etoolbox}
\usepackage{siunitx}

\makeatletter
\def\@email#1#2{%
 \endgroup
 \patchcmd{\titleblock@produce}
  {\frontmatter@RRAPformat}
  {\frontmatter@RRAPformat{\produce@RRAP{*#1\href{mailto:#2}{#2}}}\frontmatter@RRAPformat}
  {}{}
}%
\makeatother
\begin{document}

\preprint{AIP/123-QED}

\title[Short period InGaAs/AlInAs THz quantum cascade lasers]{Short period InGaAs/AlInAs THz quantum cascade laser in thin double metal cavities operating up to 188 K}

% Force line breaks with \\
\author{Sebastian Gloor}
    \email{gloorse@phys.ethz.ch}
    \affiliation{Institute for Quantum Electronics, Department of Physics, ETH Z\"urich, 8093 Z\"urich, Switzerland}%
\author{David Stark}
    \affiliation{Institute for Quantum Electronics, Department of Physics, ETH Z\"urich, 8093 Z\"urich, Switzerland}%
\author{Martin Franckié}
    \affiliation{Institute for Quantum Electronics, Department of Physics, ETH Z\"urich, 8093 Z\"urich, Switzerland}%
\author{Urban Senica}
    \affiliation{Institute for Quantum Electronics, Department of Physics, ETH Z\"urich, 8093 Z\"urich, Switzerland}%
    \affiliation{John A. Paulson School of Engineering and Applied Sciences, Harvard University, 9 Oxford Street, Cambridge, Massachusetts 02138, USA}%
\author{Mattias Beck}
    \affiliation{Institute for Quantum Electronics, Department of Physics, ETH Z\"urich, 8093 Z\"urich, Switzerland}%
\author{Giacomo Scalari}
    \email{gscalari@ethz.ch}
    \affiliation{Institute for Quantum Electronics, Department of Physics, ETH Z\"urich, 8093 Z\"urich, Switzerland}%
\author{Jérôme Faist}
    \affiliation{Institute for Quantum Electronics, Department of Physics, ETH Z\"urich, 8093 Z\"urich, Switzerland}%

\date{\today}% It is always \today, today,
             %  but any date may be explicitly specified

\begin{abstract}
We present a two-well terahertz (THz) quantum cascade laser designed for high temperature operation based on the InGaAs/AlInAs material system. The lighter effective mass and higher energy barriers increase the gain at high temperatures (T >\SI{150}{\kelvin}). When processed in copper-based double metal waveguides the devices show laser action up to a maximum operating temperature of \SI{188}{\kelvin} with a maximum current density of \SI{1.4}{\kilo \ampere / \cm \squared}. The low Joule heating due to reduced active region thickness and low electrical bias allows operation at 10$\%$ duty cycle up to a temperature of \SI{170}{\kelvin}.

\end{abstract}

\maketitle

%\section{Introduction}

    The terahertz (THz) spectral region, often defined between 1-10 THz, has been proposed for applications in astrophysics, telecommunications, non-invasive medical imaging and spectroscopy \cite{Leisawitz2000}. The lack of high-brilliance, coherent THz sources constitutes  a relevant  obstacle for such applications. The THz quantum cascade laser\cite{Scalari30years2024_comms_Phys,williams_terahertz_2007} (QCL) has since emerged as a high-power, compact THz emitter with the advantage of large wavelength agility (1.2-6 THz) \cite{Gao2023, Shahili2024}. The main drawback remains the maximum operating temperatures of THz QCLs, that still restrict most of these devices to cryogenic operation, although recent advances have enabled use of thermoelectric coolers with the maximum operating temperature reaching \SI{261}{K} \cite{Bosco2019, Khalatpour2021, Khalatpour2023}. Lately, our group demonstrated GaAs-based THz QCL operating in a compact,  commercial high-heat-load (HHL) thermoelectrical cooler packaging \cite{GloorNanophot2025}, unlocking out-of-the-lab applications for such devices.

    %Current Challenges in THz QCLs
    The main mechanism limiting the operating temperature for THz QCLs is the fast non-radiative relaxation from the upper laser state due to strong longitudinal-optical phonon (LO-phonon) emission present in polar III-V materials used for THz QCLs. Due to the emission energy being below the optical phonon energy for these materials, population inversion is strongly suppressed with increasing temperature as electrons can easily acquire the necessary in-plane kinetic energy to relax to a lower subband through LO-phonon emission. The optical gain at elevated temperatures thus becomes too low to overcome the large propagation losses present in double-metal THz QCLs, usually estimated around \SI{15}{\per \cm} to \SI{20}{\per \cm} for a double metal waveguide  of 12 $\mu$m thickness.

    %Current state of the art
    
    The highest operating temperatures for THz QCLs were achieved with two-well active regions to minimize the amount of electronic levels involved in transport, maximizing the carrier fraction available for light generation. In GaAs-based active regions, carrier leakage to continuum states was recognized to be a limiting factor in two-well active regions due to the high electric fields associated with short periods \cite{Albo2015}. Increasing the Al-content in the barriers to 25\% resulted in a maximum operating temperature of \SI{210}{\kelvin} enabling the use of four-stage Peltier coolers \cite{Bosco2019}. A further increase in barrier height to Al$_{0.35}$Ga$_{0.65}$As barriers and careful optimization of parasitic current channels lead to a T$_{\text{max}}$ of \SI{261}{\kelvin}, well within reach of single-stage thermoelectric coolers \cite{Khalatpour2023}. With increasing temperature, the thermal energy $k_bT$ (\SI{25}{\milli \eV} at \SI{300}{\kelvin}) approaches the LO-phonon energy of GaAs at $h\omega = $36 meV. This results in a thermal population (backfilling) of the lower lasing level coming from the extractor state. Detuning the extraction energy from the LO-phonon towards higher values suppresses thermal backfilling at the cost of a slower extraction. In two-well THz QCLs aimed for high temperature operation, the detuned extraction energy leads to a favorable tradeoff between thermal population and lower level lifetime \cite{Frankie2020, Rindert2022}.

    %Introduce influence of material systems
    THz QCLs have been mostly realized in the GaAs/Al$_\mathrm{x}$Ga$_\mathrm{1-x}$As material system. This is in large part due to practical reasons since GaAs and Al$_\mathrm{x}$Ga$_\mathrm{1-x}$As are lattice matched for all Al-fractions simplifying thick epitaxial growths needed for low waveguide losses \cite{Strupiechonski2011}. The conduction band discontinuity (CBD) can be varied with the Al-content in the barriers giving an additional design parameter. For deeply confined states a low CBD is desirable to minimize interface roughness scattering provided over-the-barrier leakages are not significant at the nominal operating point. Materials limiting phonon-interactions such as Si/SiGe \cite{Lynch2002}, Ge/SiGe \cite{Stark2021} and GaN/AlGaN \cite{Terashima2011} have been proposed to improve the maximum operating temperature of THz QCLs. Electroluminescence was demonstrated in these materials but no laser action could be achieved. Another way of increasing the intersubband gain is by utilizing a material with a low effective mass m$^*$, as intersubband quantum efficiency is predicted to scale as $\left(m^*\right)^{-\frac{3}{2}}$, which was verified experimentally by Beneviste et al\cite{faist_quantum_2013, Benveniste2008}. In$_{0.53}$Ga$_{0.47}$As/In$_{0.52}$Al$_{0.48}$As lattice matched to InP features a lighter effective mass of m$^*_\mathrm{InGaAs}$ = 0.0427 compared to m$^*_\mathrm{GaAs}$ = 0.068, theoretically increasing intersubband gain by a factor of $\left(\frac{m^*_\mathrm{GaAs}}{m^*_\mathrm{InGaAs}}\right)^{-\frac{3}{2}} \approx$ 2. So far InP-based THz QCL have not reached the same performance as their GaAs counterparts. A possible reason is the large CBD of \SI{520}{\milli \eV} resulting in thin barriers making structures more sensitive to growth variations. But substitutions with alternative barrier materials such as GaAsSb \cite{Deutsch2012} and InAlGaAs \cite{Ohtani2013, Ohtani2016} could not improve performance over AlInAs barriers. In this work we are using In$_{0.53}$Ga$_{0.47}$As/In$_{0.52}$Al$_{0.48}$As, where we expect the higher barriers to be advantageous for limiting leakage currents.

\newpage

%\section{\label{sec:level1}Design and Fabrication}
    %The initial design was done following the same procedure as in \cite{Frankie2020}.
    The initial design was found by optimizing the calculated gain with an in-house NEGF model with a Gaussian process optimization algorithm \cite{Frankie2020}. The design features the same considerations that lead to the latest improvement in operating temperature. The optical transition is kept diagonal at an oscillator strength of f = 0.4 to reduce LO-phonon emission depleting the upper level population. The extraction energy is detuned from the LO-phonon energy in In$_\mathrm{0.43}$Ga$_\mathrm{0.52}$As of \SI{32}{\milli \eV} to \SI{48}{\milli \eV} to reduce thermal backfilling at higher temperatures. Any subsequent design changes were simulated with the nextnano.NEGF package \cite{Grange2015}. The design resulting from the initial optimization was grown with 200 periods and is labeled EV2795. Two design iterations, EV3036 and EV3105, were grown. For EV3036 the injection barrier was increased from 25.255\AA{} to 29.343\AA{}, decreasing the injection coupling from $2\hbar\Omega$ = \SI{2.9}{\milli\eV} to $2\hbar\Omega$ = \SI{2.3}{\milli\eV}. The thicker injection barrier is intended to increase injection selectivity and reduce sub-threshold leakage currents resulting from the resonant alignment $|i_n\rangle \rightarrow  |l_{n+1}\rangle$. EV3105 instead was based on the same layer sequence as EV2795 but its period length reduced by 2.3\%, where the best performing devices for EV2795 were found due to the radial thickness variation in MBE growth. Calculated parasitic currents for EV3105 are increased but negative-differential regions (NDR) before reaching J$_\mathrm{max}$ are suppressed leading to a more electrically stable structure. For EV3036 and EV3105, the number of periods was reduced from 200 to 150 resulting in active regions of \SI{5.3}{\micro \metre} and \SI{5.24}{\micro \metre} thickness, respectively. The doping was also reduced from \SI{4.5e10}{\per \cm \squared} to \SI{3.9e10}{\per \cm \squared}. The doping is shifted from the center of the phonon well towards the injection barrier to reduce impurity scattering out of the upper laser level \cite{Frankie2020}. The design parameters for the different layers can be seen in Table \ref{tab:sim_overview}. The bandstructure was calculated with a Schrödinger-Poisson solver and the relevant parameters extracted at a lattice temperature of \SI{300}{\kelvin}. The oscillator strength $f_\mathrm{i,p}$ from the injector to the first parasitic level has been identified as an important parameter for leakage currents that correlates with the achieved T$_\mathrm{max}$ \cite{Khalatpour2021}. $f_\mathrm{i,p}$ is very sensitive to the bias and material parameters used in the simulation. Direct comparison is difficult because of the different material system but our values are comparatively larger than what was calculated for their devices. The gain was then simulated in nextnano.NEGF and included 7 bound states with electron-electron scattering enabled.

        \begin{table}[h]
    \caption{\label{tab:sim_overview} 
    \\
    The following quantities are extracted from Schrödinger-Poisson bandstructure calculations: Injection coupling $2\hbar\Omega_\mathrm{u,l}$, Oscillator strength between laser levels $f_\mathrm{u,l}$ and between injector and first parasitic level $f_\mathrm{i,p}$, Extraction energy $E_\mathrm{ex}$, Energy difference between injector and first parasitic state $E_\mathrm{i,p}$.
    }
    \begin{ruledtabular}
    \begin{tabular}{ccccccc}
    % Quantitiy

    % Simulation
    Layer &
    $2\hbar\Omega_\text{u,l}$ (meV) &
    $f_\text{u,l}$ &    
    $E_\text{ex}$ (meV) &
    $E_\text{i,p}$ (meV) &
    $f_\text{i,p}$ &

    \\
    \hline
    EV2795 & 2.87 & 0.4 & 48 & 45.9 & 0.33\\
    EV3036 & 2.1 & 0.42 & 47.8  & 48.8 & 0.25 \\
    EV3105 & 3.15 & 0.44 & 49.6 & 47.2 & 0.36 \\
    EV3181 & 2.9 & 0.4 & 48.3 & 45.1  &  0.375 \\
    \end{tabular}
    \end{ruledtabular}
\end{table}

    \begin{figure*}[h]
    \includegraphics[width=1\linewidth]{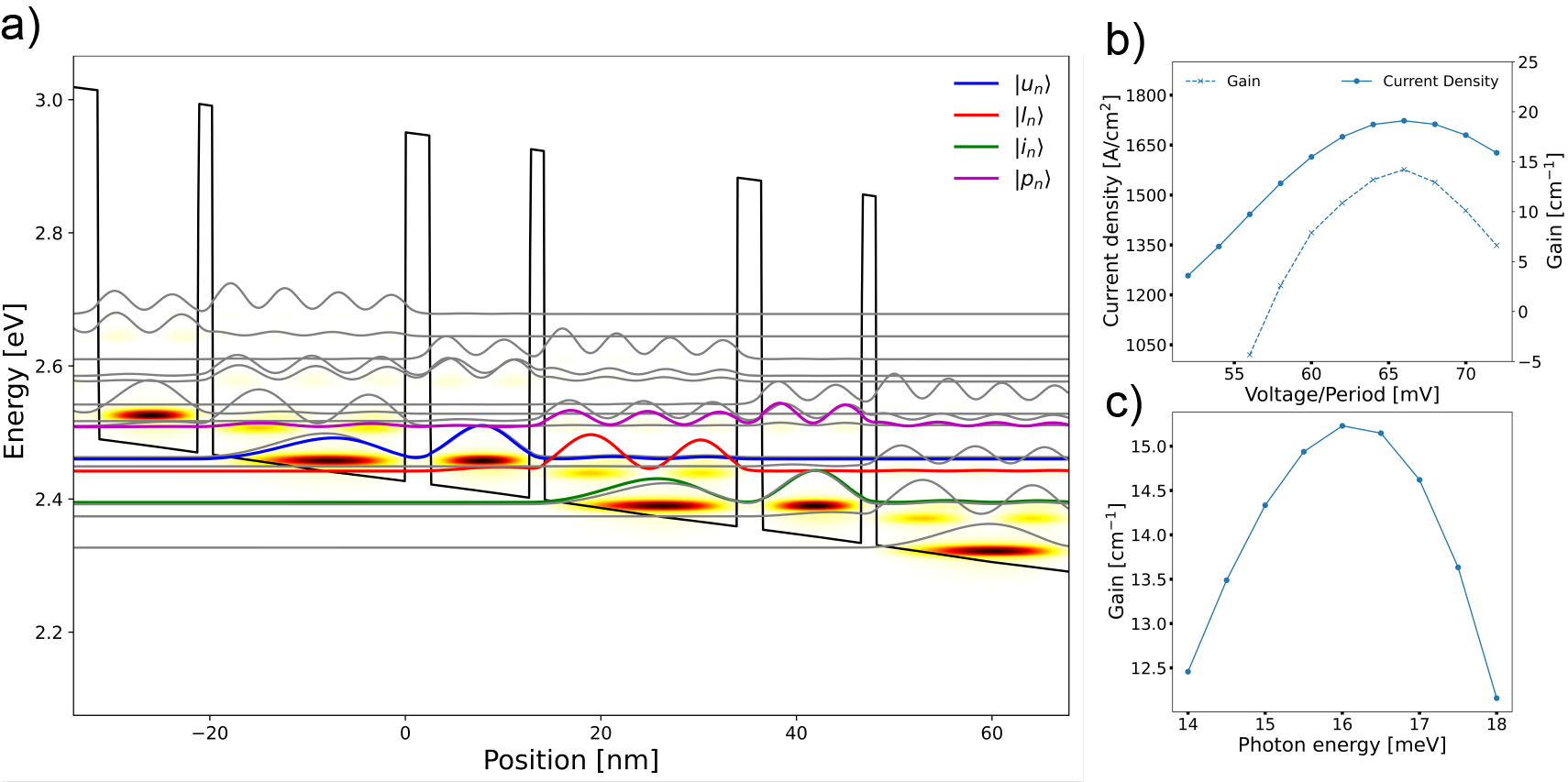}
    \caption{\label{fig:Simulation}
    \textbf{a)} Bandstructure of EV2795 in the Wannier-Stark basis. The laser levels as well as the first parasitic level are highlighted. The colorscale shows the carrier concentration as calculated by NEGF. \textbf{b)} Gain and current density in dependence of bias per period of EV2795. Calculations were performed at \SI{300}{\kelvin} with e-e scattering enabled. \textbf{c)} Calculated gain spectrum at a bias of \SI{66}{\milli \volt /period}.   
    \\}    
    \end{figure*}

    Wafers were grown with a solid-source MBE and subsequently cleaved into pieces measuring approximately \SI{12}{\milli \metre} $\times$ \SI{10}{\milli \metre}. EV2795 was processed with Ti/Au waveguides and an intact \SI{60}{\nano \metre} thick n$^+$ top contact, while EV3036 and EV3105 were processed with Ta/Cu waveguides and the highly doped layer chemically removed before metal deposition. Laser ridges were defined by chemical etching in a 1:1:1 H$_3$PO$_4$:H$_2$O$_2$:H$_2$O solution. EV3105 was subsequently also processed with ICP-RIE using a Cl$_2$/H$_2$ chemistry. This achieves the vertical sidewalls that are needed for the best performance. Ridges of \SI{750}{\micro \metre} to \SI{1.75}{\milli \metre} lengths were cleaved and In-soldered onto copper submounts.

%\section{\label{ExpRes}Experimental results}
    Devices were characterised with a He-cooled Si-Bolometer and electrically driven with \SI{50}{\nano \s}- \SI{70}{\nano \s} long pulses at \SI{415}{\hertz} to match the bolometer's response. The spectrum could be measured under vacuum with a step scan interferometer in the same setup.

    The first growth of the nominal design, EV2795, was grown on a full 3" InP-wafer and a radial study could be performed. From XRD-measurements a period scaling from +2.5\% to -2.5\% is found from center to the edge of the wafer. A maximum operating temperature of \SI{115}{\kelvin} was measured for devices from the edge with center pieces reaching \SI{100}{\kelvin}. The maximum current density also increased by 77\% from center to the edge of the wafer. These devices showed very limited dynamic range even for the best devices of $\eta = (J_{max}-J_{th})/J_{max} = $ 0.12. The laser frequency shifts from center to edge from \SI{13.8}{\milli \eV} to \SI{16}{\milli \eV}. Devices from this epilayer showed significant electrical instabilities near $J_\mathrm{max}$ with an early-occuring negative differential resistance (NDR) limiting the laser performance \cite{Fathololoumi2013}. The stability of devices was improved for shorter period lengths where a higher T$_\mathrm{max}$ was reached. The voltage bias at J$_\mathrm{max}$ is lower than what would be expected from NEGF simulations further suggesting an early occurring NDR.

\begin{figure*}[h]
    \includegraphics[width=1\linewidth]{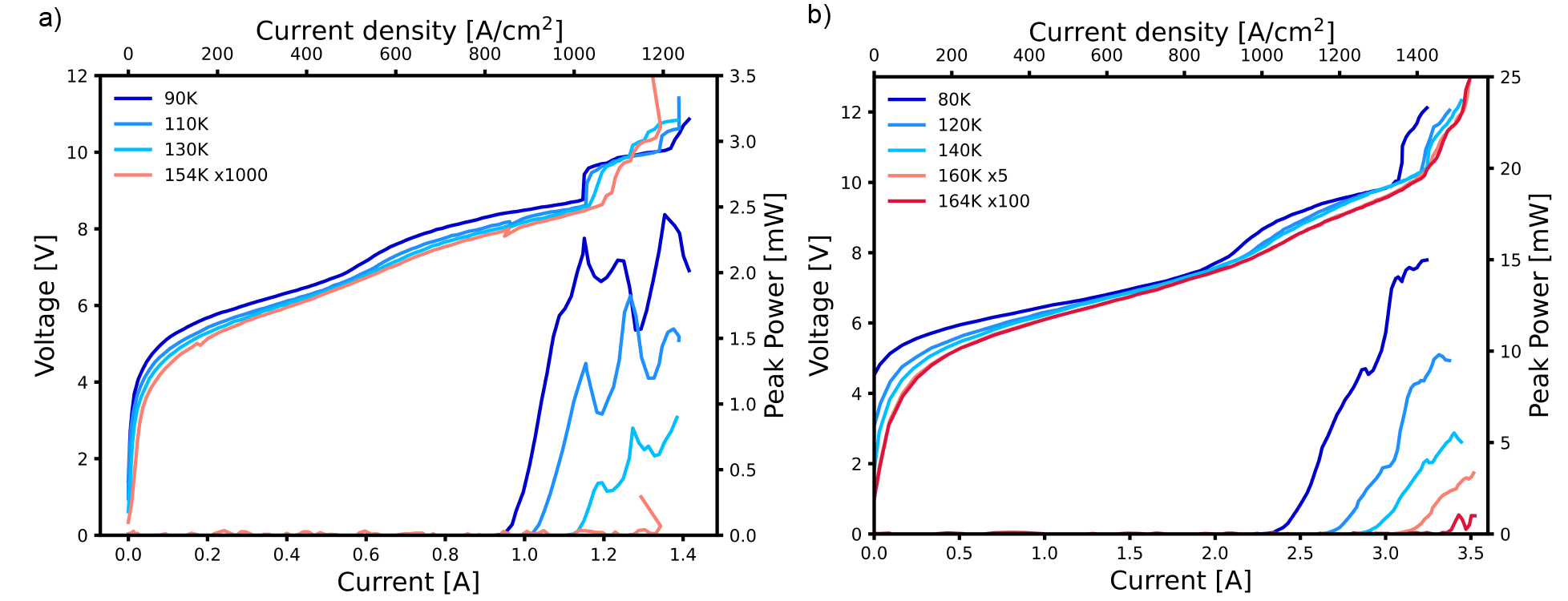}
    \caption{\label{fig:EV3036}
    \textbf{a)} Light-Current-Voltage characterisation of a wet etched ridge device from epilayer EV3036. The layer sequence is \textbf{29.34}/102.181/\textbf{14.721}/99.865/\underline{30}/67.348 with AlInAs barriers in bold and the underlined layer doped with \SI{1.3}{\per \cm \cubed}. The device is \SI{750}{\micro \metre} long and \SI{150}{\micro \metre} wide. Maximum operating temperature is \SI{154}{\kelvin}. \textbf{b)} Light-Current-Voltage characterisation of a wet etched, \SI{1.75}{\milli \metre} long and \SI{130}{\micro metre} wide ridge device from epilayer EV3105 lasing up to \SI{164}{\kelvin}. The nominal layer thicknesses are \textbf{25.255}/102.181/\textbf{14.721}/99.865/\underline{30}/67.348 with the same doping as EV3036. XRD measurements determined the period length to be -2.3\% of the nominal value.
    }
    \end{figure*}

    Measurements for devices from EV3036 are reported in Fig. \ref{fig:EV3036} a). A maximum operating temperature of \SI{154}{\kelvin} was measured. The impact of the thicker injection barrier can clearly be seen by the shift of the alignment resonance to lower currents compared to EV3105. The dynamic range was significantly improved to $\eta = $ 0.35 although an increase is also expected from lower waveguide losses resulting in lower threshold current densities. At a bias of around \SI{8.6}{\volt} there is a distinct jump in the IV-curve that is not seen in devices with Ti/Au waveguides of the same epilayer. This behaviour was reproduced in the majority of devices and is due to a spontaneous rearrangement of the field domains inside the structure \cite{Fathololoumi2013}. There is no significant change in frequency before and after the jump. Between \SI{130}{\kelvin} and \SI{154}{\kelvin} $J_\mathrm{max}$ decreases from \SI{1230}{\ampere \per \cm \squared} to \SI{1194}{\ampere \per \cm \squared} suggesting a lifetime limited transport regime \cite{Khalatpour2023}. The electrical instabilities have not been completely eliminated with this design.

\begin{figure*}
    \includegraphics[width=1\linewidth]{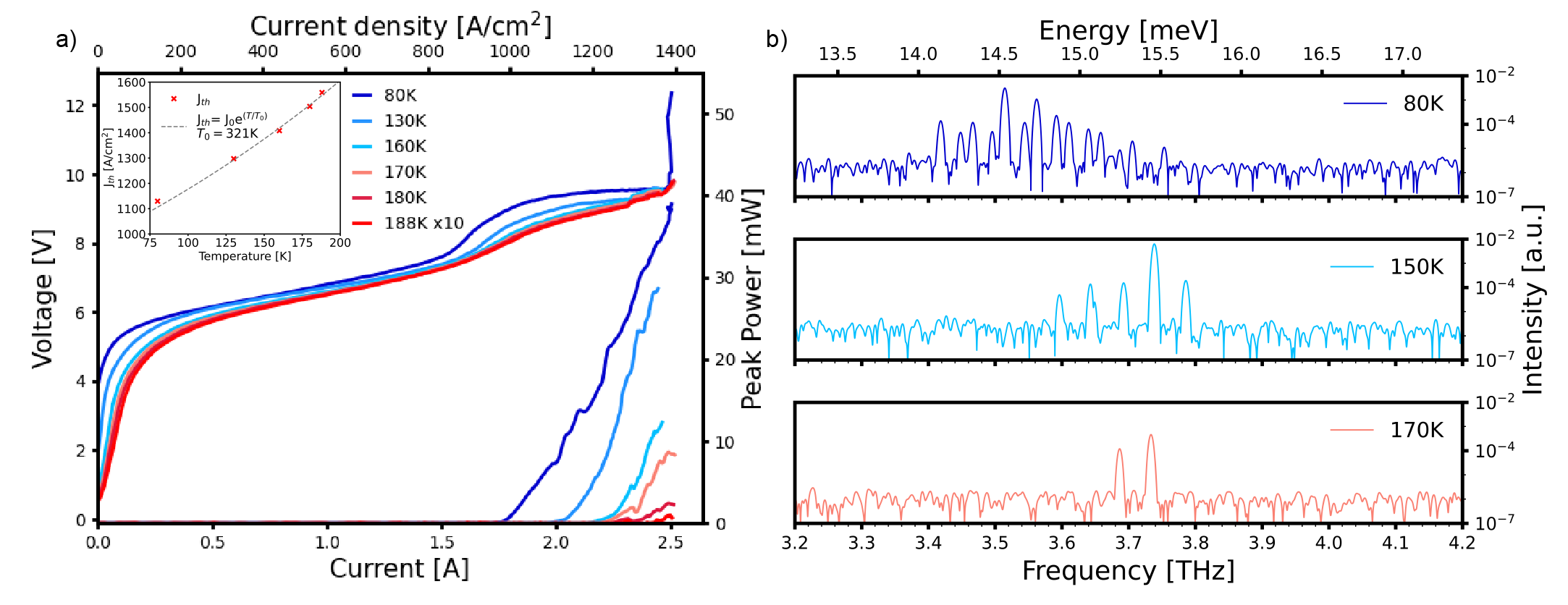}
    \caption{\label{fig:EV3105_LIV}
    \textbf{a)} Light-Current-Voltage characterisation of EV3105. The device is \SI{1.2}{\milli \metre} long and \SI{150}{\micro \metre} wide. The laser was driven with $\sim$ 70ns long pulses at a repetition rate of \SI{415}{\Hz} and lases up to \SI{188}{\kelvin}. \textbf{Inset:} J$_\mathrm{th}$ behaviour with increasing temperature. The best device shows exceptionally high T$_0$=\SI{344}{\kelvin}. \textbf{b)} Spectral emission for a \SI{750}{\micro \metre} long and \SI{100}{\micro \metre} wide ridge device driven at 20\% duty cycle. At lower temperature the laser operates in a broadband regime due to the highly diagonal lasing transition collapsing to a narrowband state around \SI{3.7}{\tera \hertz} at temperatures above \SI{170}{\kelvin}.
    }
    \end{figure*}

    EV3105 on the other hand again showed the resonant alignment of injector and lower laser level close to laser threshold with a dynamic range of $\eta = $ 0.24. The increase of threshold current density with temperature can be empirically described with $J_\mathrm{th} = J_0 \exp(\frac{T}{T_0})$. EV3105 shows a high T$_0$ = \SI{261}{\kelvin} compared to EV3036 T$_0$ = \SI{220}{\kelvin}. The maximum operating temperature was increased to \SI{164}{\kelvin} improving over previous highest operating temperature of InGaAs/AlInAs THz QCLs by \SI{10}{\kelvin}. The emission frequency closely matches devices processed from the edge of wafer EV2795, showing the reproducibility of the growth. EV3105 was then reprocessed in a dry-etch to reach the highest operating temperature for this layer by achieving vertical sidewall geometry. Measurements can be seen in Fig. \ref{fig:EV3105_LIV}. The best performing device lased up to a heatsink temperature of \SI{188}{\kelvin}. Devices from this epilayer show high T$_0$ values with the best performing device exhibiting a T$_0$ of \SI{321}{\kelvin} although super exponential increase is usually observed for the last measurement points. The slope efficiency of these devices remains constant between \SI{80}{\kelvin} and \SI{170}{\kelvin}. From a rate equation approach we find $\frac{dP}{dI} = \frac{N_ph\nu}{e}\frac{\alpha_m^{\mathrm{front}}}{\alpha_{\mathrm{tot}}}\frac{\tau _{\mathrm{eff}}}{\tau _{\mathrm{eff}} + \tau _2}$. The constant slope efficiency suggests a favorable increase in $\frac{\tau _{\mathrm{eff}}}{\tau _{\mathrm{eff}} + \tau _2}$ with temperature without significant thermally activated carrier leakage channels. In Fig. \ref{fig:EV3105_LIV} b) spectral measurements on a commercial rapid scan FTIR for a \SI{750}{\micro \metre} long and \SI{130}{\micro \metre} wide ridge driven at 20\% duty cycle is shown. At \SI{80}{\kelvin} the laser shows fairly broadband emission of around \SI{350}{\giga \hertz} centered around \SI{3.55}{\tera \hertz} owing to the broadened, diagonal optical transition. The spectrum blue shifts with increasing temperature and emits around \SI{3.7}{\tera \hertz} at \SI{170}{\kelvin}. Such a high duty cycle only \SI{12}{\kelvin} below the T$_\mathrm{max} =$  \SI{182}{\kelvin} for this device can be achieved due to the low electrical dissipation for a two-well structure. With a bias of only \SI{10}{\volt} and \SI{1.4}{\kilo \ampere \per \cm \squared} the dissipation is a factor of four lower than what is usually recorded in GaAs-based two-well active regions \cite{Bosco2019}. 
    While these devices already constitute a significant improvement over previous InGaAs/AlInAs THz QCLs in operating temperature, it remains to be determined as to why the performance of GaAs-based two-well THz QCLs could not be reached. The high CBD of \SI{520}{\milli \eV} results in increased interface roughness and gain broadening. The estimation of these parameters is difficult resulting in less accurate predictions from simulations. The presented epilayers are only \SI{5.3}{\micro \metre} thick, significantly increasing waveguide losses due to larger overlap with the metals. It has been shown that a reduction from \SI{12}{\micro \metre} to \SI{5}{\micro \metre} can result in a change in T$_\mathrm{max}$ of \SI{20}{\kelvin} \cite{Strupiechonski2011}. Double metal waveguide losses are usually placed in the range of \SI{10}{\per \cm} to \SI{15}{\per \cm} for \SI{12}{\micro \metre} thick active regions. The waveguide losses scale roughly inversely linear in ridge height which places our losses between \SI{20}{\per \cm} to \SI{30}{\per \cm}. This coincides with predicted gain from NEGF at the measured T$_\mathrm{max}$. The low current density of our active regions are beneficial from a thermal standpoint but less useful for achieving the highest T$_\mathrm{max}$ and higher doping levels could lead to higher operating temperatures, considering that we are likely limited by waveguide losses.

\newpage

%\section{Conclusions}
We presented two-well InGaAs/AlInAs THz QCLs, leveraging the high conduction band offset to limit over-the-barrier leakage currents and the lighter effective mass for increased intersubband gain. Devices lase up to \SI{188}{\kelvin}, improving the maximum operating temperature by \SI{33}{\kelvin} over previous InGaAs/AlInAs THz QCLs. The devices do not reach yet the performance of GaAs/AlGaAs devices possibly limited by increased interface roughess scattering, lower doping levels or the relatively thin active regions at \SI{5.3}{\micro \metre} resulting in higher waveguide losses. The electrical dissipation in these devices is very low due to the absence of a contact effect at the InGaAs/metal interface. This allows for higher duty cycles and average output powers at temperatures close to T$_\mathrm{max}$ and would be an important feature if the maximum temperature reaches the cooling floor of thermoelectric coolers \cite{GloorNanophot2025}.

%\section{Acknowledgements}
The author acknowledges Filippo Ciabattini for his advice in fabricating the samples and Adrian Weisenhorn for support of the measurements and the cleanroom facility FIRST of ETH Zürich.
This work was part of 23FUN03 COMOMET that has received funding from the European Partnership on Metrology, co-financed by the European Union’s Horizon Europe Research and Innovation Programme and from by the Participating States. Funder ID: 10.13039/100019599. We gratefully acknowledge funding received from the Swiss National Science Foundation (SNF grant No. 200021-232335).

%\section{Author Declarations}
The authors have no conflicts of interest to disclose.
%\section{Data Availability}
The datasets generated and/or analysed during the current study are not publicly available but are available from the corresponding author on reasonable request.

\newpage

\nocite{*}

\bibliography{References}% Produces the bibliography via BibTeX.
%\bibliography{References}
\end{document}